# Significant anisotropic neuronal refractory period plasticity

RONI VARDI[2], YAEL TUGENDHAFT[1], SHIRA SARDI[1] and IDO KANTER[1,2,*]

[1]Department of Physics, Bar-Ilan University, Ramat-Gan, 52900, Israel.
[2]Gonda Interdisciplinary Brain Research Center and the Goodman Faculty of Life Sciences, Bar-Ilan University, Ramat-Gan, 52900, Israel.
[*]e-mail: ido.kanter@biu.ac.il

**Abstract -** Refractory periods are an unavoidable feature of excitable elements, resulting in necessary time-lags for re-excitation. Herein, we measure neuronal absolute refractory periods (ARPs) in synaptic blocked neuronal cultures. In so doing, we show that their duration can be significantly extended by dozens of milliseconds using preceding evoked spikes generated by extracellular stimulations. The ARP increases with the frequency of preceding stimulations, and saturates at the intermittent phase of the neuronal response latency, where a short relative refractory period might appear. Nevertheless, preceding stimulations via a different extracellular route does not affect the ARP. It is also found to be independent of preceding intracellular stimulations. All these features strongly suggest that the anisotropic ARPs originate in neuronal dendrites. The results demonstrate the fast and significant plasticity of the neuronal ARP, depending on the firing activity of its connecting neurons, which is expected to affect network dynamics.

**Introduction.** – Refractory periods (RPs) are an unavoidable feature of excitable elements, given that following excitation, the formation of necessary conditions for a re-excitation demands a recovery period. This universal feature governs the activity of excitable elements ranging from stimulated and spontaneous emission from atoms[1] or spiking chaotic lasers[2] to the repetition of epileptic seizures[3]. Similarly, an unavoidable feature of the neuron, the excitable element of our brain, is the absolute refractory period (ARP), which entails a certain degree of repulsion between consecutive evoked spikes[4-7]. It stems from the time-lag required for internal processes to generate a new evoked spike. Traditionally, the ARP was associated with the neuronal membrane hyperpolarization after an evoked spike[7-9] and it was estimated to last a few milliseconds. Recent experimental results on synaptic blocked neuronal cultures indicate that the phenomenon of ARP exists without visible neuronal hyperpolarization[10,11], and that its duration can exceed 20 milliseconds, which is comparable to that of the neuronal membrane decay timescale[12,13]. In any event, the ARP phenomenon is assumed to be an imprint time-independent intrinsic property of the neuron, which is independent of the activity of its connecting neurons. This long-lasting assumption[14-16] about the fundamental principles governing neuronal behavior is hereby challenged.

In this study, we present the following three main features of the ARP, measured in synaptic blocked neuronal cultures[10,11]. First, ARP is not an imprint time-independent property of the neuron. Its duration depends on the history of the neuron's preceding spiking activity, and its extended reversible duration can exceed a factor eight. The second feature is the anisotropic property of the ARP. Preceding spiking activity generated from one stimulation route, e.g. a dendrite, does not affect the duration of the neuronal ARP measured at another neuronal route, e.g. a different dendrite. This anisotropic phenomenon[11,17] strongly suggests that the mechanism generating the ARP has to be associated with a region outside of the soma, and, likely, with the dendrites[18,19]. This is enhanced by the third feature whereby preceding spiking activity generated by intracellular stimulations does not affect the ARP duration.

**Results.** – Our experimental setup consisted of neuronal cultures plated on a multi-electrode-array combined with a patch clamp, with added synaptic blockers[10,11] (see Methods). To measure the ARPs, pairs of extracellular stimulations, with monotonically increasing intra-pair time-lags (fig. 1(a)), were given to a patched neuron. The stimulating extracellular electrode, as well as the amplitude of the stimulations, is preselected to ensure reliable responses at low-frequency stimulations. In addition, pairs of stimulations were separated by a time-lag $\tau$, which is much greater than the intra-pair time-lag (fig. 1(a)), to ensure a non-overlapping effect between consecutive pairs. To measure the effect of preceding spiking activity on the ARP, the same stimulation scheduling was enforced with additional $N$ preceding extracellular stimulations given at a higher frequency, $f_h$ (fig. 1(b)).

The preceding stimulations result in an increase of the neuronal response latency (NRL) by several milliseconds[20-24], as measured by the time-lag between the extracellular stimulation and its corresponding evoked spike (fig. 1(c)). The NRL increases with the stimulation frequency[25] and is saturated at the intermittent phase for long enough sequence of stimulations given at a frequency greater than the neuron's critical frequency[25], $f_c$ (fig. 1(c)). The intermittent phase is characterized by fluctuations of the NRL around a constant value and with average maximal firing frequency, $f_c$ (fig. 1(C)), which varies much among neurons[25]. This maximal firing frequency, $f_c$, is independent of the stimulation frequency wherein $f_h > f_c$, and is regulated by neuronal response failures (fig. 1(c)); e.g. for $f_h = 2f_c$, an average 50% of the stimulations result in response failures.

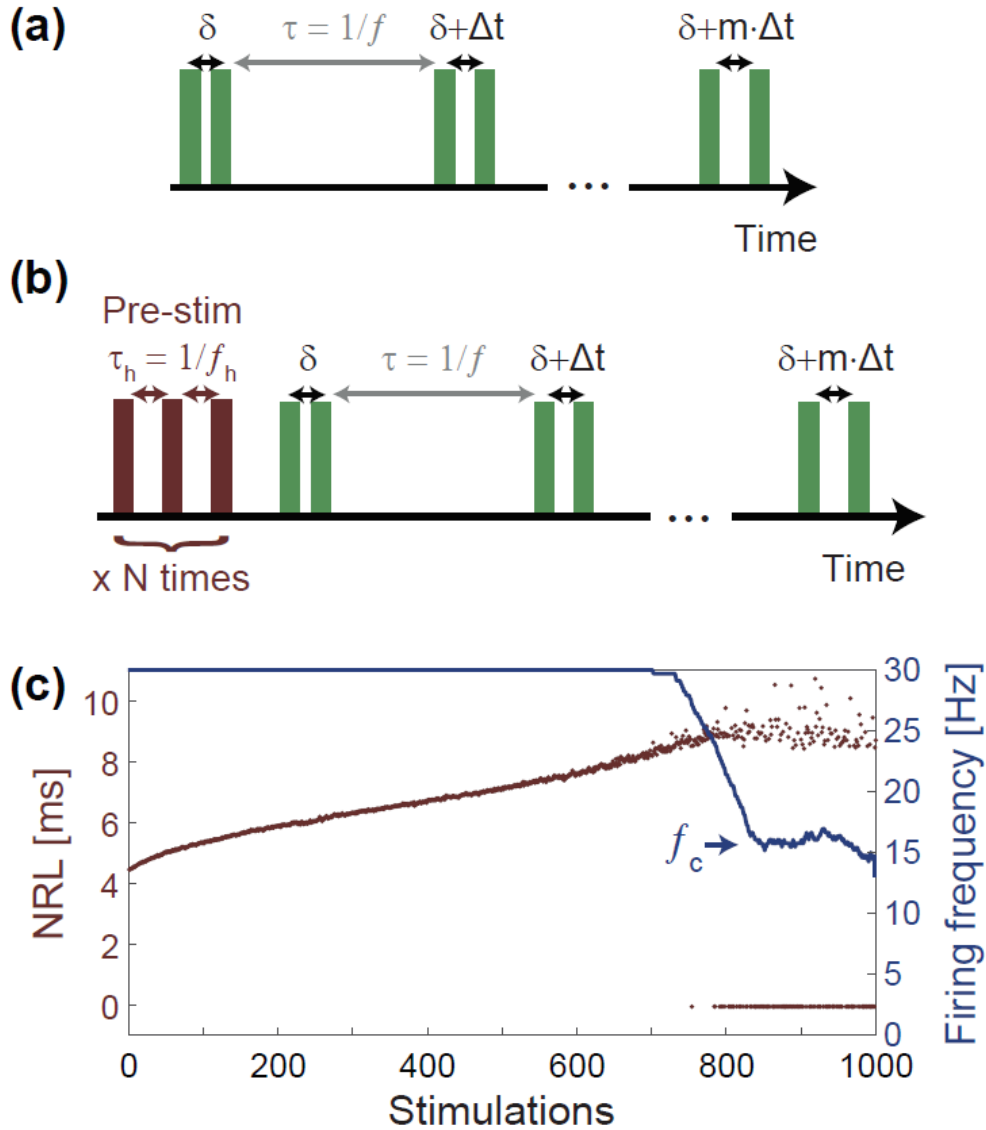

Fig. 1: (color online). Extracellular stimulation scheduling for measuring the neuronal refractory period. (a) Pairs of extracellular stimulations given at a fixed frequency, $f$, with monotonically increasing internal time-lags, $\delta + m \cdot \Delta t$, where, $\delta$ is the first time-lag and $m$ is the number of pairs. (b) The same scheduling as in (a) with additional N preceding extracellular stimulations (brown) given at a fixed frequency $f_h$. (c) Experimental results of a neuron in a blocked culture, stimulated 1000 times extracellularly with $f_h = 30$ Hz. The neuronal response latency, NRL, measuring the time-lag between the extracellular stimulations and its corresponding intracellular measured evoked spike (brown) and the firing frequency is averaged over a sliding window of 100 stimulations (blue). Response failures are denoted at zero NRL. The average firing frequency at the intermittent phase is denoted by $f_c$ ~15 Hz.

The first type of experimental setup consists of a patched neuron within a synaptic blocked culture, which is stimulated via an extracellular electrode (fig. 2(a)). The ARP is defined as the shortest time-lag between pair of stimulations which result in response to both stimulations (first appearance of two green "v"). The RRP is defined as the shortest time-lag between pair of stimulations which no response-failures were recorded after it (last appearance of red "x"). The stimulation scheduling (fig. 1(a)) indicates a sharp transition to a full responsiveness[7,8] of the second stimulation within each pair for an intra-pair time-lag $\geq 4.5\ ms$ (fig. 2(b), left), which is around the intracellular measured 5 $ms$ ARP (fig. 2(b), right). Note that there is a small difference, typically less than 1.5 ms, between the time-lag of intra-pair stimulations and the time-lag between corresponding spikes. This difference is a result of small fluctuations in the NRL. This sharp transition from the ARP to a fully responsiveness occurs without a relative refractory period (RRP). Using 1000 preceding stimulations at $f_h$ =7 Hz for the same neuron, the NRL increased by ~2 $ms$ and the ARP increased from 5 $ms$ to 9 $ms$, without RRP (fig. 2(c)), (Supplementary figure S5, n=9, mean=4.1 ms, std=1.4 ms). Note that the intrinsic variations of the stimulations are below a millisecond (Supplementary figure S7).

Further increasing the preceding stimulation frequency of the same neuron to $f_h = 12\ Hz$ saturates the fluctuating NRL around $10\ ms$, where the neuron enters the intermittent phase after ~700 stimulations which is characterized by $f_c \sim 10\ Hz$ (fig. 2(d)). In this case, the ARP increased to $40\ ms$, followed by a short RRP; a full responsiveness is observed above $45.5\ ms$ intra-pair time-lag. The ARP increases by about a factor of eight, from $5\ ms$ to $40\ ms$, while the neuron's NRL is in the intermittent phase (Supplementary figure S5, n=9, mean=23.8 ms, std=9.9 ms). The increase in ARPs by the preceding stimulation frequency, which is enhanced while entering the intermittent phase, was found to be consistent in our experiments and appears to reflect a general phenomenon. This increase (fig. 2(d) and Supplementary figure S5) cannot be attributed to the response failures[26] at the intermittent phase[21] (fig. 1(d)), since no response failures are observed in the first stimulation of each pair (fig. 2(d)).

The appearance of RRP in the intermittent phase is avoidable and depends on the stimulation frequency of the pairs, $f$ (fig. 1(b)). Typically, the RRP appears where $2f \gg f_c$ and disappears for $2f \ll f_c$, where $2f$ is the average neuronal stimulation frequency by the pairs (fig. 2(e)). In addition, the ARP also typically increases at the intermittent phase for $2f \gg f_c$ (fig. 2(e)). Within this limit, response failures can occur for the first, the second, or both stimulations within a pair (Supplementary figure S1). Nevertheless, after the intermittent phase was achieved, an increase in the ARP was observed even at low-frequency pairs, $2f \ll f_c$, where the NRL decays from the intermittent phase (Supplementary figure S2), pointing to the possible long timescales of this phenomenon.

The second type of experiment consists of the stimulation of the patched neuron using two different extracellular electrodes, (purple and green in fig. 3(a)), generating different spike waveforms (fig. 3(a), right), i.e. stimulating the patched neuron via two distinct routes. The preceding stimulations can be carried out either via the green electrode or the purple electrode; however, the ARP is always measured using pairs of stimulations (fig. 1(a)) via the green electrode. The ARP without preceding stimulations ended when the time-lag between intra-pair stimulations was $6.5\ ms$ (fig. 3(b)). Preceding stimulations at a high frequency, $f_h = 10$ Hz, via the purple electrode result in the NRL intermittent phase with $f_c \sim 3.3\ Hz$ (fig. 3(c), right), although a moderate ARP increased to $8.5\ ms$, and without RRP (fig. 3(c), left). In contrast, preceding stimulations at a frequency of $f_h = 15$ Hz via the green electrode result in the NRL intermittent phase with $f_c \sim 8\ Hz$ (fig. 3(d), right) and $14.5\ ms$ ARP, and a short RRP (fig. 3(d), left). The neuronal ARP is anisotropic, whereby preceding stimulation sequences occur via two routes, generating different spike waveforms with different NRL profiles and $f_c$, thereby resulting in two different durations of ARPs. Limited experimental results (n=3) indicate that the ARPs increases after preceding stimulations at a high frequency via the green electrode (mean = 9.6 ms), but almost no increase was found after preceding stimulations at a high frequency via the purple electrode (mean = 1 ms). This effect is consistent with fluctuations of less than 1 ms in repeated measurements of the ARP in the same neuron (Supplementary figure S7).

The third type of experiments consists of preceding intracellular stimulations (similar to fig. 1(b)), followed by pairs of intracellular stimulation scheduling (similar to fig. 1(a)). This setup consists of precise timings of intracellular stimulating and recording of the patched neuron (fig. 4($a_1$)). The ARP was found to be the same, $5.5\ ms$, without/with preceding intracellular stimulations (fig. 4($a_2$) and fig. 4($a_3$), respectively). Supplementary figure S6 indicates that for n=14, the intrinsic variations of the intracellular stimulations are below half a millisecond. The same neuron but with pairs of *extracellular* stimulating scheduling (fig. 1(a)) results in $7\ ms$ ARP (fig. 4($b_1$)); however, with preceding extracellular stimulations (fig. 1(b)) leading to NRL intermittent phase, the ARP increased to $16\ ms$ (fig. 4($b_2$)). Results strongly indicate that the intracellularly measured ARP is not affected by preceding intracellular stimulation (fig. 4(a)) or by preceding extracellular stimulations, as indicated by preliminary results (Supplementary figure S4). The extracellular measure of the ARP using high frequency intracellular preceding stimulations warrants a new type of experiment, which, at the moment, is beyond our experimental setup's capacity.

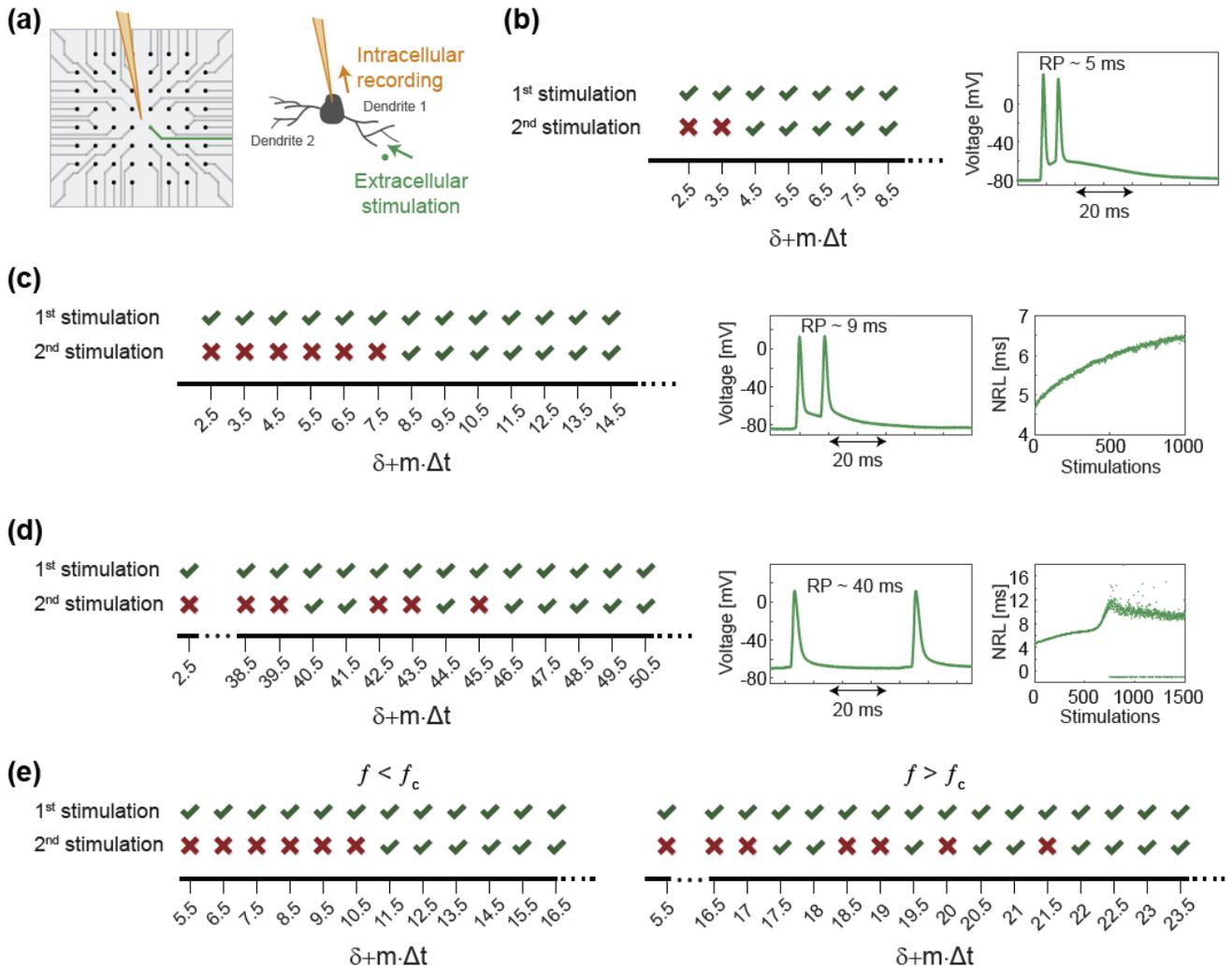

Fig. 2: (color online). Measuring the neuronal refractory period without/with preceding extracellular stimulations. (a) Schematic representation of the experimental setup of a multi-electrode-array (left) whereby a patched neuron in a blocked culture is stimulated via an extracellular electrode (green) and recorded intracellularly (orange). (b) The stimulation scheduling (fig. 1(a)) with $\delta = 1.5\ ms$, $\Delta = 1\ ms$ and $f = 2\ Hz$, resulting in $5\ ms$ ARP. Evoked spike/response failure for each pair of stimulations is denoted by green-v/brown-x, respectively (left). The intracellular recorded voltage of the first pair of evoked spikes, separated by $4.5\ ms$, is shown at the right panel. (c) The same neuron in (b) with $N = 1000$ preceding stimulations, $f_h = 7\ Hz$ and $f = 0.5\ Hz$ (fig. 1(b)), the ARP increased from $5\ ms$ to $9\ ms$ (middle), and the NRL did not enter the intermittent phase (right). (d) The same neuron in (c) with $N = 1500$ preceding stimulations, $f_h = 12\ Hz$ and $f = 6\ Hz$ (fig. 1(b)), where the NRL entered the intermittent phase (right). ARP $40\ ms$ and a short RRP appears. (e) A different neuron where preceding stimulations results in the intermittent phase, $f_c \sim 7\ Hz$ (Supplementary figure S3). Left: Consecutive pairs of stimulations are given at $f = 2\ Hz$ ($2f < f_c$) (fig. 1(b)), resulting in $11.5\ ms$ ARP and no RRP. Right: Consecutive pairs of stimulations are given at $f = 4\ Hz$ ($2f > f_c$), resulting in $17.5\ ms$ ARP followed by RRP terminating at $22.5\ ms$.

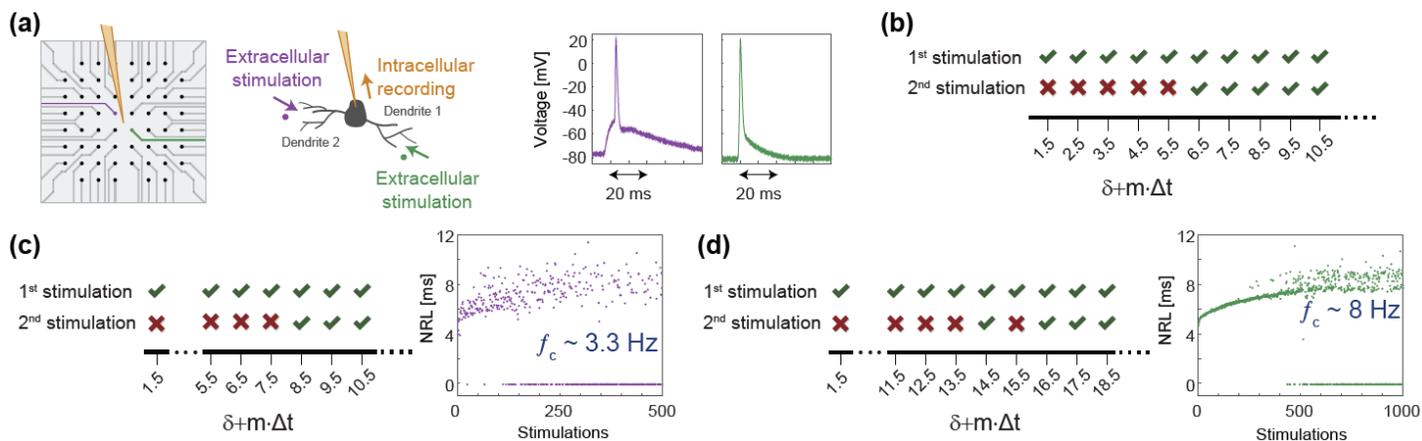

Fig. 3: (color online). Measuring the neuronal refractory period without/with preceding stimulations from different extracellular electrodes. (a) Schematic representation of the experimental setup of a multi-electrode-array (left) in which a patched neuron in a blocked culture is stimulated via two extracellular electrodes (green and purple) generating two different spike waveforms (right), and recorded intracellularly (orange). (b) The stimulation scheduling (fig. 1(a)) using the green electrode with δ=1.5 ms, Δ=1 ms and f=0.5 Hz. The ARP without preceding stimulations ended when the time-lag between intra-pair stimulations was 6.5 ms. (c) The same neuron in B with preceding stimulations via the purple electrode, N=500 and $f_h\sim$ 10 Hz, where the NRL entered the intermittent phase (right), followed by pairs of stimulations via the green electrode, with f=0.5 Hz, resulting in 8.5 ms ARP. (d) Similar to (c), but with preceding stimulations, N=1000 and $f_h\sim$ 15 Hz are provided via the green electrode, where the NRL entered the intermittent phase (right), followed by pairs of stimulations via the same green electrode, with f=0.5 Hz, resulting in 14.5 ms ARP, and with RRP ending at 16.5 ms.

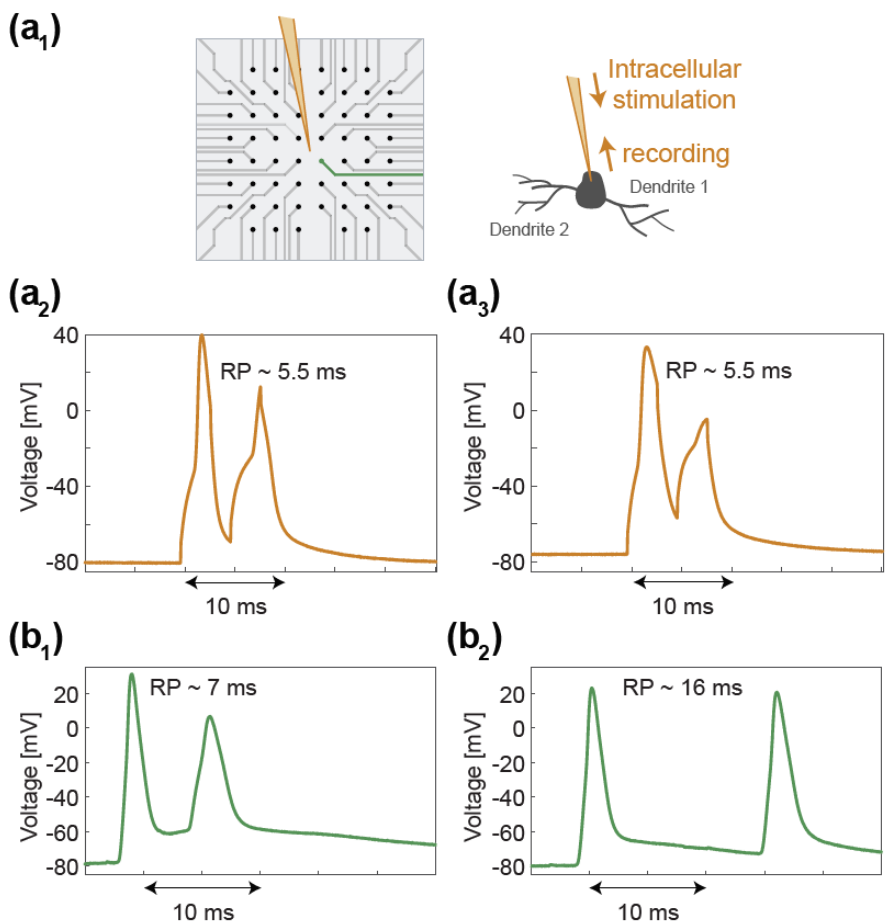

Fig. 4: (color online). Measuring the neuronal refractory period without/with preceding intracellular stimulations. ($a_1$) Schematic representation of the experimental setup of a multi-electrode-array (left) in which a patched neuron in a blocked culture is stimulated and recorded via an intracellular electrode (orange). ($a_2$) Pairs of intracellular stimulations (similar to fig. 1(a)) with δ=1 ms, Δ=1 ms and f=2 Hz, resulting in 5.5 ms ARP. ($a_3$) Similar to A2 with N=500 intracellular preceding stimulations, and $f_h$=30 Hz (similar to fig. 1(b)), resulting in the same ARP, 5.5 ms. ($b_1$) The same neuron in (a), stimulated by pairs of extracellular stimulations (similar to fig. 2(b)), resulting in 7 ms ARP. ($b_2$) Similar to ($b_1$), with preceding extracellular stimulations, N=800 and $f_h$=15 Hz, resulting in 16 ms ARP.

**Discussion.** – The anisotropic features of the ARP (fig. 3) and its independent duration with preceding intracellular stimulations (fig. 4) strongly indicate that the mechanism regulating its duration is located out of the soma, most likely in the dendrites. This is also consistent with previous results indicating that the neuron functions as multiple threshold units, [11] with anisotropic transmission rates[27] and as an anisotropic adaptive node[17]. Some of these anisotropic features are irreversible, such as the neuronal different spike waveforms, whereas others, such as the ARP and NRL, are reversible[25]. The question of whether all these reversible neuronal anisotropic features are governed by the same timescale remains unclear and deserves further advanced experiments. An answer might shed light on their underlining mechanisms.

The extended duration of the ARP using preceding stimulations varies among neurons. In some neurons, the ARP is extended by a few milliseconds only, while in others it is extended by a factor eight (fig. 2(d)). Nevertheless, one rule governs all our measured neurons: The ARP increases with increasing NRL and its maximal duration is achieved at the intermittent phase. However, a priori prediction of the maximal duration of the ARP based on the knowledge of the spike waveform, $f_c$ and the profile of the NRL, remains unknown. In addition, the distribution of the maximal ARP extension among neurons using preceding stimulation remains unknown. Our experiments suggest, for instance, that neurons with a small NRL increase, e.g. by 1–2 ms, or with a concave profile of the NRL increase, typically result in moderate extension of the ARP. Quantitatively answering these questions warrants further experiments.

**Methods.** – The In-Vitro experimental methods are similar to our previous studies[11,17] and only the modifications are presented.
*Animals Use,* All procedures were in accordance with the. National Institutes of Health Guide for the Care and Use of Laboratory Animals and Bar-Ilan University Guidelines for the Use and Care of Laboratory Animals in Research and were approved and supervised by the Bar-Ilan University Animal Care and Use Committee.
*Neuronal response latency*, The neuronal response latency (NRL) is defined as the time-lag between a stimulation pulse onset and its corresponding evoked spike measured by crossing a threshold of -10 mV. *Statistical analysis,* Reported results are based on 50 examined cultures and different examined neurons, where each type of experiment was repeated at least 5 times.
*Data analysis,* Analyses were performed in a MATLAB environment (MathWorks, Natwick, MA, USA). The recorded data from the MEA (voltage) was filtered by convolution with a Gaussian that has a standard deviation (STD) of 0.1 ms. Evoked spikes were detected by threshold crossing, −10 mV, using a detection window of 0.5−20 ms following the beginning of an electrical stimulation.

REFERENCES

[1] A. Friedman, A. Gover, G. Kurizki, S. Ruschin, and A. Yariv, Reviews of modern physics **60**, 471 (1988).
[2] M. Rosenbluh, Y. Aviad, E. Cohen, L. Khaykovich, W. Kinzel, E. Kopelowitz, P. Yoskovits, and I. Kanter, Physical Review E **76**, 046207 (2007).
[3] D. K. Naritoku, D. J. Casebeer, and O. Darbin, Epilepsia **44**, 912 (2003).
[4] A. L. Hodgkin and A. F. Huxley, The Journal of physiology **117**, 500 (1952).
[5] P. R. Gray, Biophysical Journal **7**, 759 (1967).
[6] M. A. Brazier, Science, 1423 (1964).
[7] M. J. Berry II and M. Meister, in *Advances in neural information processing systems*1998), pp. 110.
[8] S. Kuffler, R. Fitzhugh, and H. Barlow, The Journal of general physiology **40**, 683 (1957).
[9] P. Kara, P. Reinagel, and R. C. Reid, Neuron **27**, 635 (2000).

[10]	W. Moolenaar and I. Spector, The Journal of physiology **292**, 297 (1979).
[11]	S. Sardi, R. Vardi, A. Sheinin, A. Goldental, and I. Kanter, Sci Rep-Uk **7**, 1 (2017).
[12]	V. Shul'govskii, A. Moskvitin, and B. Kotlyar, Neurophysiology **7**, 359 (1975).
[13]	F. R. Fernandez, J. Noueihed, and J. A. White, Journal of Neuroscience **39**, 2221 (2019).
[14]	H. Nakahama and S. Nishioka, J Theor Biol **12**, 140 (1966).
[15]	R. B. Stein, Biophysical Journal **5**, 173 (1965).
[16]	W. R. Amberson, The Journal of physiology **69**, 60 (1930).
[17]	S. Sardi, R. Vardi, A. Goldental, A. Sheinin, H. Uzan, and I. Kanter, Sci Rep-Uk **8**, 1 (2018).
[18]	N. Spruston, Nature Reviews Neuroscience **9**, 206 (2008).
[19]	M. London and M. Häusser, Annu. Rev. Neurosci. **28**, 503 (2005).
[20]	G. Aston-Jones, M. Segal, and F. E. Bloom, Brain Res **195**, 215 (1980).
[21]	R. De Col, K. Messlinger, and R. W. Carr, The Journal of physiology **586**, 1089 (2008).
[22]	J. van Pelt, P. S. Wolters, M. A. Corner, W. L. Rutten, and G. J. Ramakers, IEEE Transactions on Biomedical Engineering **51**, 2051 (2004).
[23]	D. J. Bakkum, Z. C. Chao, and S. M. Potter, Plos One **3**, e2088 (2008).
[24]	E. V. Evarts, Science **179**, 501 (1973).
[25]	R. Vardi, A. Goldental, H. Marmari, H. Brama, E. A. Stern, S. Sardi, P. Sabo, and I. Kanter, Front Neural Circuit **9**, 29 (2015).
[26]	N. Spruston, Y. Schiller, G. Stuart, and B. Sakmann, Science **268**, 297 (1995).
[27]	R. Vardi, A. Goldental, A. Sheinin, S. Sardi, and I. Kanter, EPL (Europhysics Letters) **118**, 46002 (2017).

# Supplementary Material

## Significant anisotropic neuronal refractory period plasticity

RONI VARDI[2], YAEL TUGENDHAFT[1], SHIRA SARDI[1] and IDO KANTER[1,2,*]

[1]*Department of Physics, Bar-Ilan University, Ramat-Gan, 52900, Israel.*
[2]*Gonda Interdisciplinary Brain Research Center and the Goodman Faculty of Life Sciences, Bar-Ilan University, Ramat-Gan, 52900, Israel..*
[*]*e-mail:* [ido.kanter@biu.ac.il](ido.kanter@biu.ac.il)

(a)

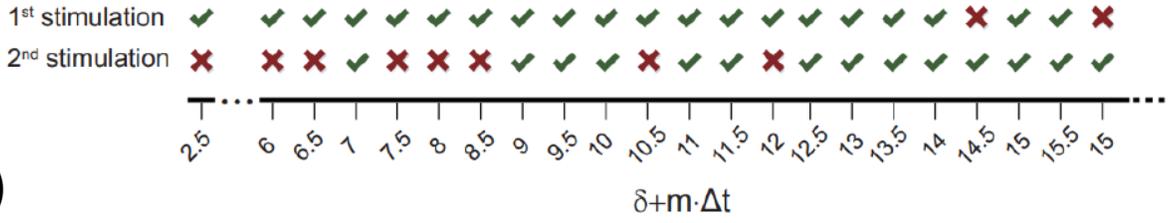

(b)

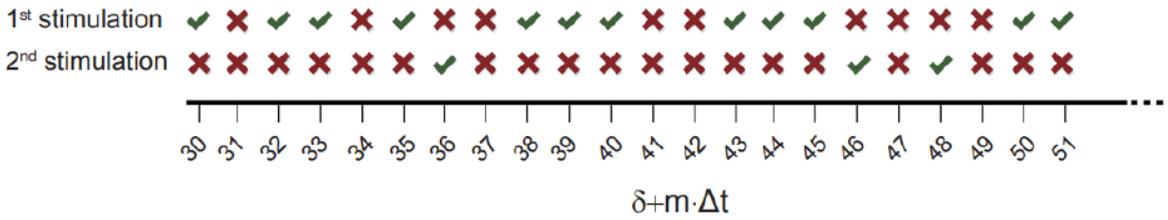

Figure S1: Response failures can occur for the first, the second or both stimulations within a pair while the NRL is in the intermittent phase. (a) The stimulation scheduling with δ=2.5 ms, Δ=0.5 ms, f=6 Hz and $f_c$=~5.5 Hz, with preceding stimulations, N=200, $f_h$=15 Hz (Figure 1(b)). Evoked spike/response failure for each pair of stimulations is denoted by green-v/brown-x, respectively. Response failures appear for the first or second stimulations within a pair. (b) Similar to (a), with a different neuron and δ=30 ms, Δ=1 ms, f=7.5 Hz, $f_c$=~5 Hz, N=800, $f_h$=15 Hz. Response failures appear for the first, second, or both stimulations within a pair.

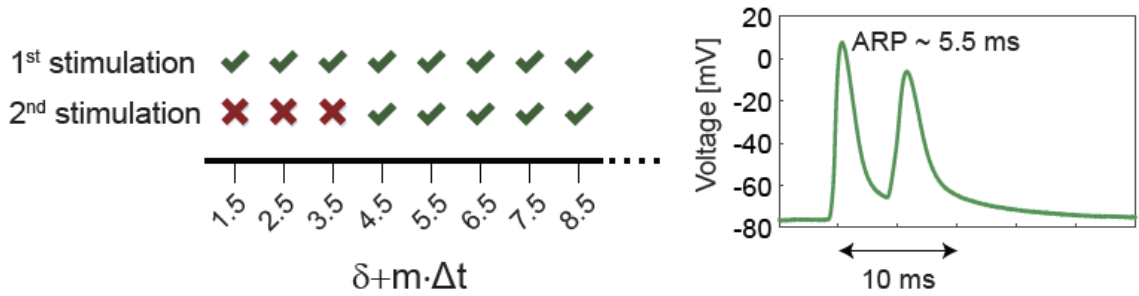

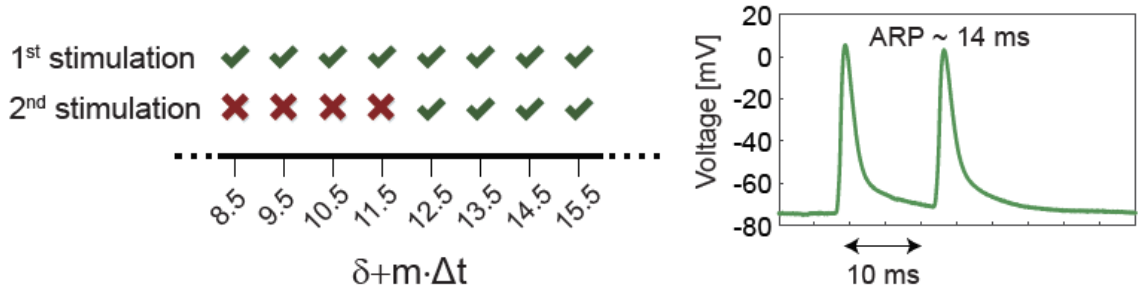

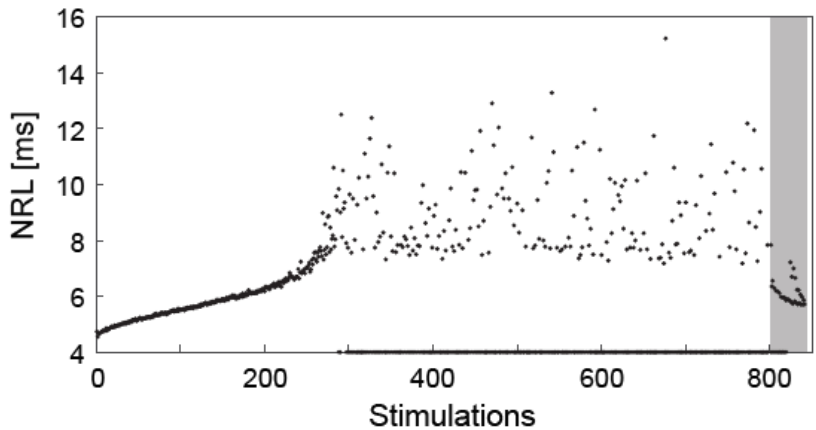

Figure S2: A significant increase in the ARP while the NRL decays from the intermittent phase, indicating longer timescales of ARP increase. (a) The stimulation scheduling (fig. 1(a)) with δ=1.5 ms, Δ=1 ms and f=2 Hz, resulting in ARP ~5 ms. Evoked spike/response failure for each pair of stimulations is denoted by green-v/brown-x, respectively (left). The intracellular recorded voltage of the first pair of evoked spikes, separated by 5.5 ms, is shown at the right panel. (b$_1$) The same neuron in A with preceding stimulations (fig. 1(b)) and low-frequency pairs, 2f≪ f$_c$~ 8 Hz, N=800, f$_h$=25 Hz and f=2 Hz (left), the ARP measured between evoked spikes increased from ~5 ms to ~14 ms (right). (b$_2$) The NRL of (b$_1$), measuring the time-lag between the extracellular stimulations and its corresponding intracellular measured evoked spike, where response failures are denoted at 4 ms NRL. After the intermittent phase was achieved, the NRL significantly decayed from the intermittent phase (gray bar).

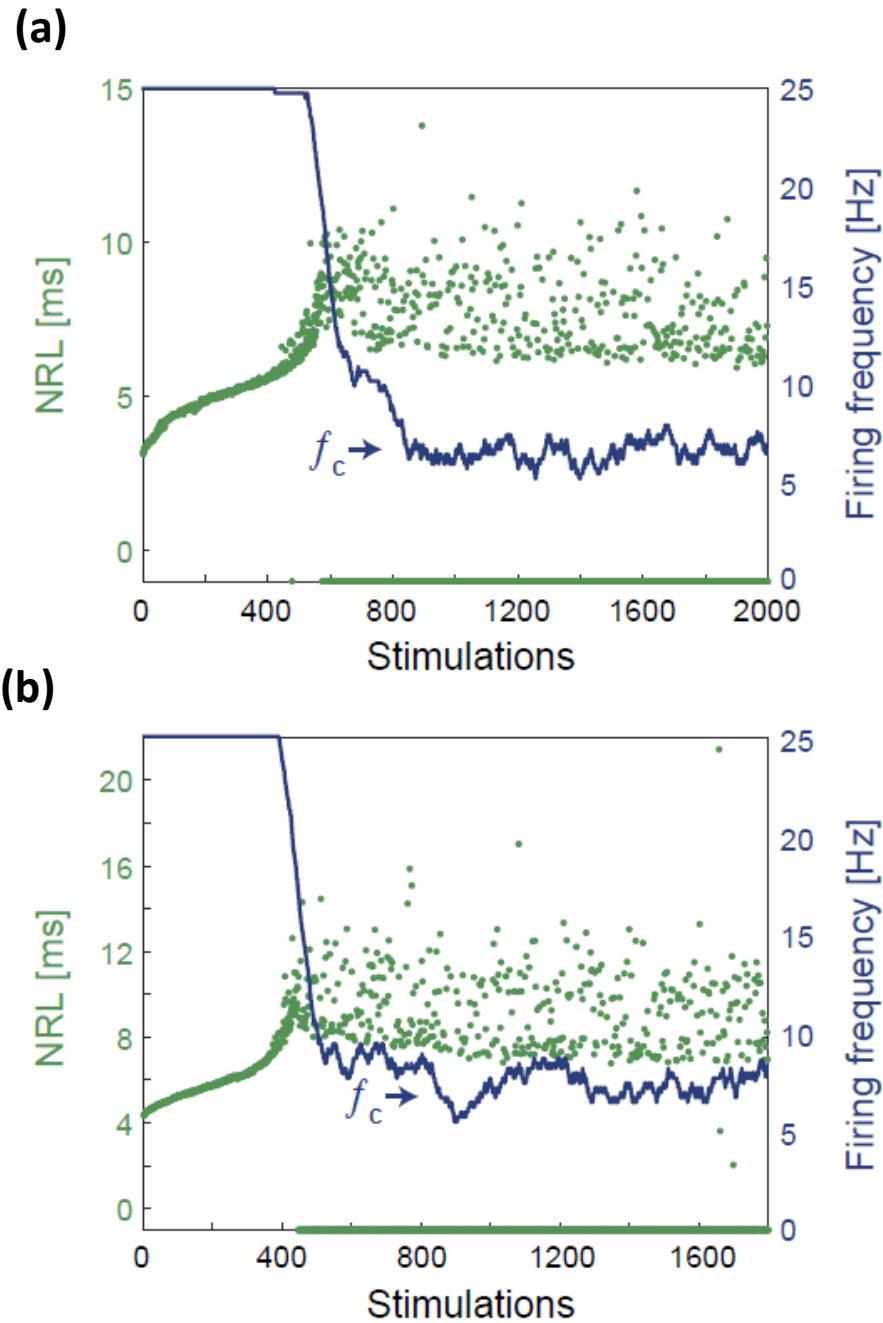

Figure S3: Experimental results of the neuron presented in figure 2(e) (main text), where preceding stimulations results in the intermittent phase. (a) Preceding stimulations (fig. 1(a)) of figure 2(e) right panel, where $N = 2000$, $f_h = 25$ Hz. The NRL, measuring the time-lag between the extracellular stimulations and its corresponding intracellular measured evoked spike (green) and the firing frequency is averaged over a sliding window of 100 stimulations (blue). Response failures are denoted at NRL= -1. The average firing frequency at the intermittent phase is denoted by $f_c \sim 7$ Hz. (b) Similar to (a) for figure 2(e) left panel, where $N = 1800$, $f_h = 25$ Hz, $f_c \sim 7$ Hz.

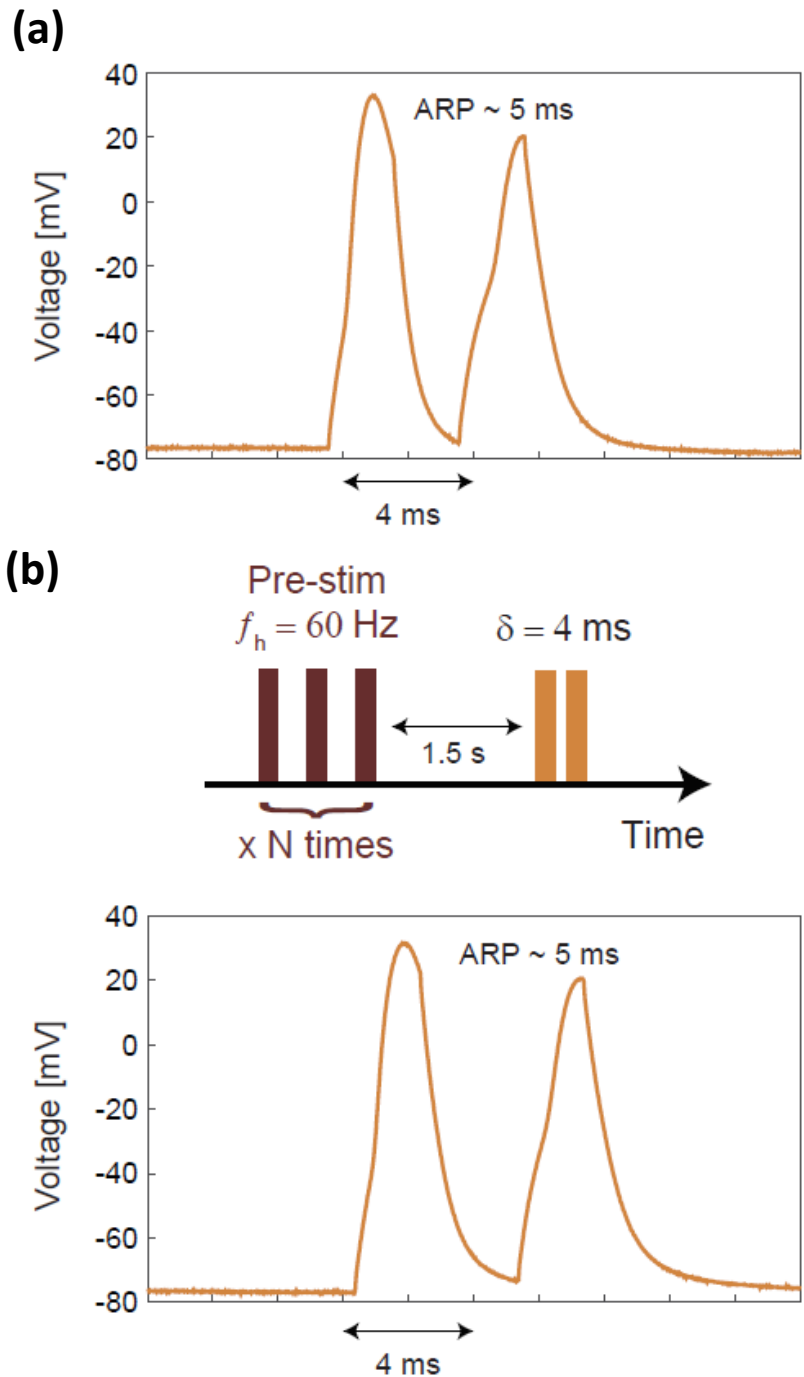

Figure S4: The intracellularly measured ARP is not affected by preceding extracellular stimulations. (a) Pairs of intracellular stimulations (similar to figure 1(a)) with $\delta = 1$ ms, $\Delta = 1$ ms and $f = 5$ Hz, resulting in ARP ~5 ms. (b) Upper panel: Similar to figure 1(b), stimulation scheduling where N=900 preceding extracellular stimulations (brown) were given at a frequency $f_h = 60$ Hz, followed by pairs of intracellular stimulations given at a fixed frequency, $f = 5$ Hz, with monotonically increasing internal time-lags, $\delta + m \cdot \Delta t$, where $\delta = 2$ ms, $\Delta = 1$ ms. Lower panel: Experimental results of the scheduling presented at the top of this panel, where pairs of intracellular stimulations (similar to (a)) result in ARP ~5 ms.

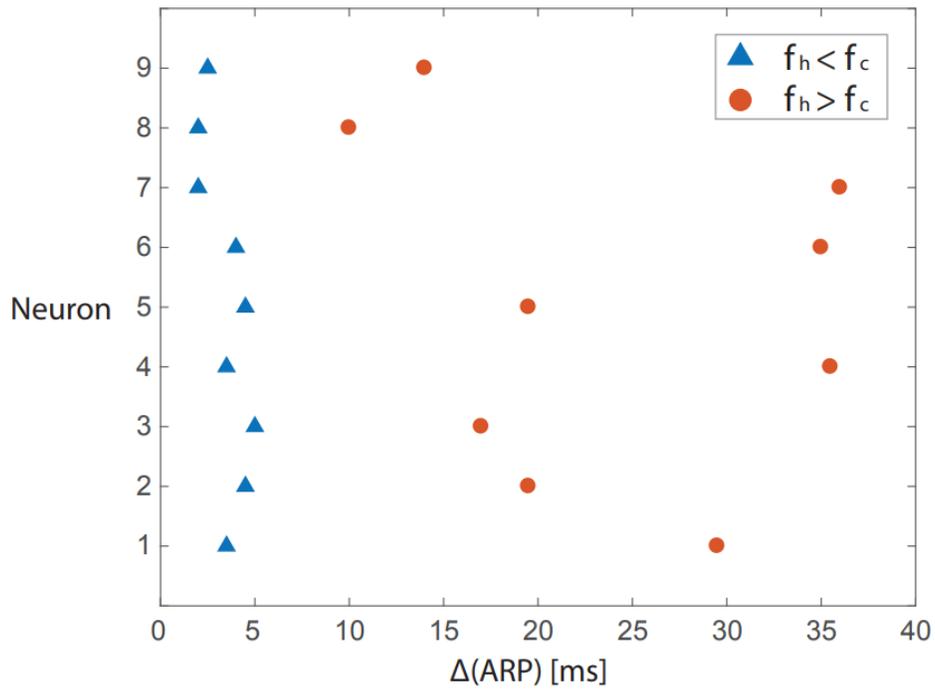

Figure S5: Measuring the neuronal refractory period without/with preceding extracellular stimulations (figure 2 in the manuscript). Measurements of the increase in the ARP of nine neurons after preceding extracellular stimulations at frequencies above (orange dots, mean=23.8 ms, std=9.9 ms) and below (blue triangles, mean=4.1 ms, std= 1.4 ms) $f_c$, in comparison to the ARP without preceding extracellular stimulations.

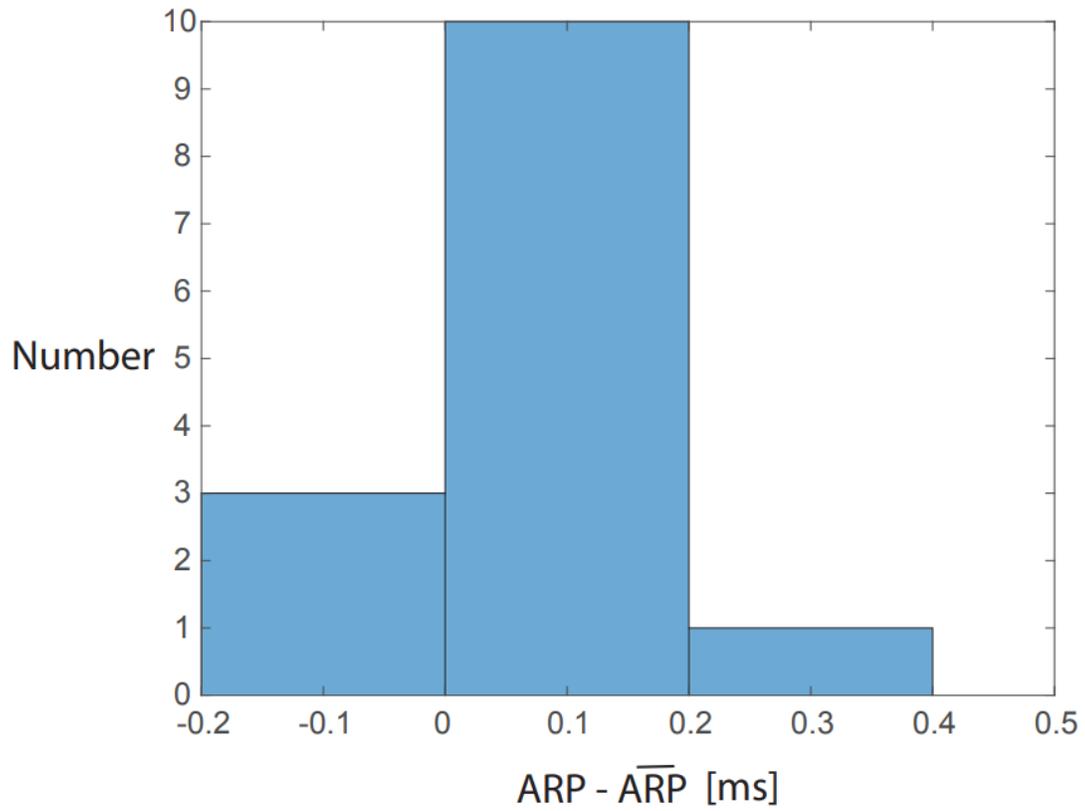

Figure S6: Measuring the neuronal refractory period without/with preceding intracellular stimulations (figure 4 in the manuscript). A histogram of the deviations of repeated measurements of the ARP with/without preceding intracellular stimulations in n=14 measurements of four neurons from their averaged value. Results indicate fluctuations below half a millisecond.

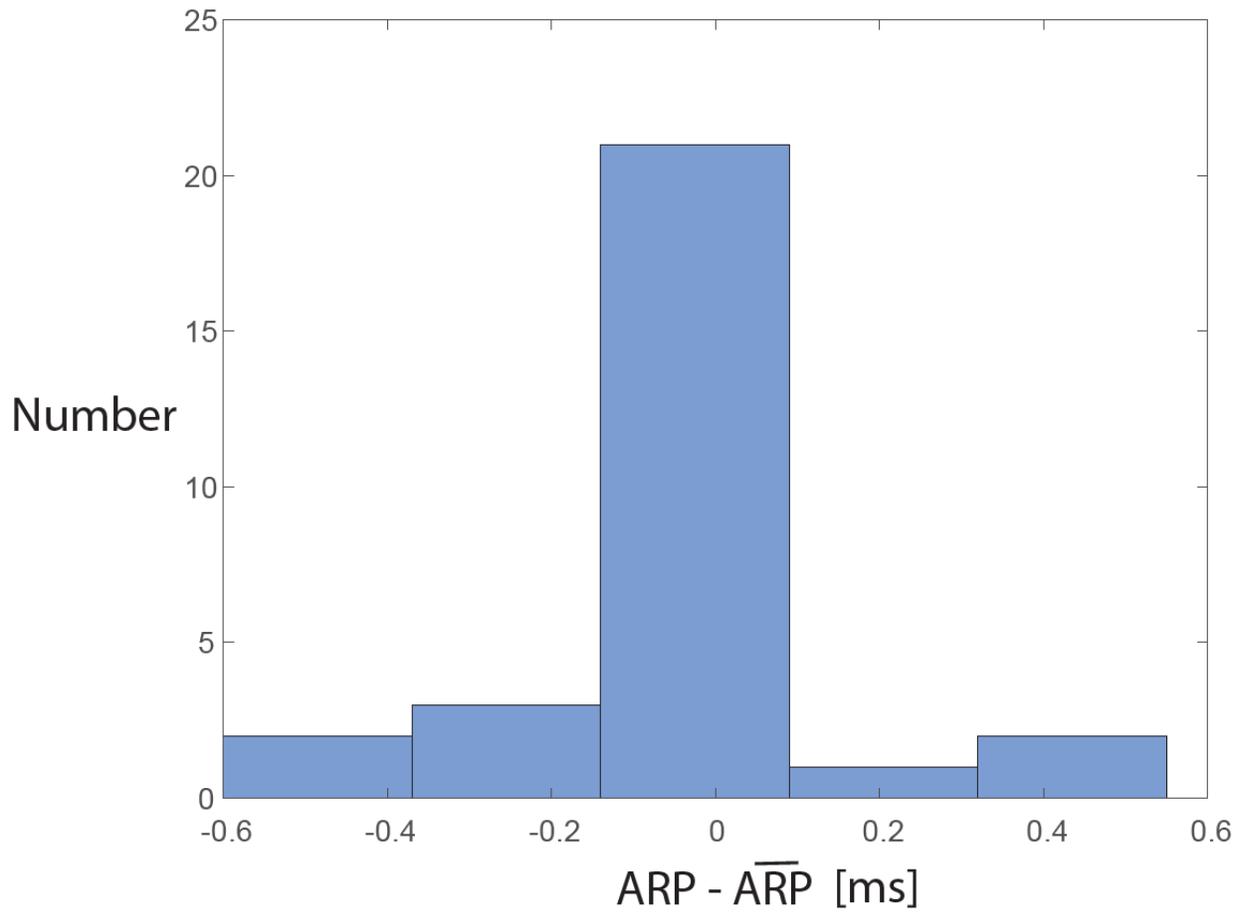

Figure S7: Repeated measurements of the neuronal refractory period. A histogram of the deviations of repeated measurements of the ARP in each of eight neurons from their averaged value. Results indicate fluctuations below a millisecond.